\documentclass[a4paper]{article}

\usepackage{INTERSPEECH2020,url}
\usepackage{multirow}
\usepackage{xcolor}

\usepackage[colorlinks = true,
            linkcolor = blue,
            urlcolor  = blue,
            citecolor = blue,
            anchorcolor = blue]{hyperref}
            
\title{Domain-Dependent Speaker Diarization for the Third DIHARD Challenge}

\name{A Kishore Kumar$^1$, Shefali Waldekar$^1$, Goutam Saha$^1$, Md Sahidullah$^2$}
\address{
  $^1$Dept. of Electronics and ECE, Indian Institute of Technology Kharagpur,
Kharagpur, India\\
  $^2$Universit\'{e} de Lorraine, CNRS, Inria, LORIA, F-54000, Nancy, France}
\email{}

\begin{document}

\maketitle
\begin{abstract}
This report presents the system developed by the ABSP Laboratory team for the third DIHARD speech diarization challenge. Our main contribution in this work is to develop a simple and efficient solution for acoustic domain dependent speech diarization. We explore speaker embeddings for \emph{acoustic domain identification} (ADI) task. Our study reveals that i-vector based method achieves considerably better performance than x-vector based approach in the third DIHARD challenge dataset. Next, we integrate the ADI module with the diarization framework. The performance substantially improved over that of the baseline when we optimized the thresholds for agglomerative hierarchical clustering and the parameters for dimensionality reduction during scoring for individual acoustic domains. We achieved a relative improvement of $9.63\%$ and $10.64\%$ in DER for core and full conditions, respectively, for Track 1 of the DIHARD III evaluation set.
\end{abstract}
\noindent\textbf{Index Terms}: acoustic domain identification (ADI), agglomerative hierarchical clustering (AHC), diarization error rate (DER), Jaccard error rate (JER), speaker diarization.

\section{Introduction}
\emph{Speaker diarization} is the task of generating time-stamps with respect to the speaker labels~\cite{Anguera2012}. For \emph{rich transcription} (RT) of spoken documents involving multiple speakers, besides ``what is being spoken" and ``by whom", it is also imperative to know ``who spoke when". In earlier studies, RT researchers mainly considered audio recordings of broadcast news, telephone conversations and meetings/conferences~\cite{Anguera2012}. With the rise in multimedia content over the years, more variety is observed in the recording environments of the spoken documents\footnote{\url{https://dihardchallenge.github.io/dihard3/index}}. Thus, instead of a \emph{one size fits all} method, a domain-dependent speaker diarization approach might cater to the diverse and challenging recording conditions in a better way.    

In our submission to the \textbf{Track 1} of the third DIHARD challenge~\cite{ryant2020third}, we propose a simple but efficient method for acoustic domain identification (ADI) using speaker embeddings. Next, we show that domain-dependent threshold for speaker clustering helps to improve the diarization performance only when probabilistic linear discriminant analysis (PLDA) adaptation uses audio-data from all the domains. Additionally, experimental optimization of the principal component analysis (PCA) parameters in a domain-specific way further improves the diarization performance.

\section{The third DIHARD challenge}
The third edition of the DIHARD challenge series, DIHARD III, has two tracks: \textbf{Track 1}: Diarization using reference SAD and \textbf{Track 2}: Diarization from scratch. The challenge data set is taken from eleven diverse domains. This brings a lot of variability in data conditions like background noise, language, source, number of speakers, and speaker demographics. The database is divided into development and evaluation sets which consists of 5 - 10 minute duration samples. Evaluation has to be done on two partitions of the evaluation data: Core evaluation set - a balanced set where total duration of the each domain is approximately equal, and Full evaluation set - where all the samples are considered for each domain. Primary evaluation metric of the challenge is \emph{diarization error rate} (DER), which is the sum of all the errors associated with the diarization task. \emph{Jaccard error rate} (JER) is the secondary evaluation metric which is based on the Jaccard similarity index. The details of the challenge with rules are available in the challenge evaluation plan~\cite{ryant2020third}.

\begin{figure}[t]
\centering
\includegraphics[scale=0.4]{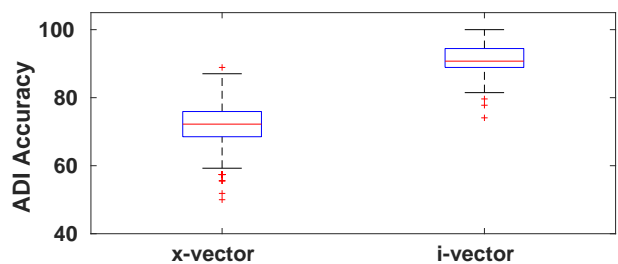}
\caption{Acoustic domain identification performance using x-vector and i-vector embeddings.}
\vspace{-0.7cm}
\label{Fig:ADIaccuracy}
\end{figure}

\begin{table*}[h]
  \caption{Results showing the impact of domain-dependent processing on speaker diarization performance (DER in \%/JER in \%) for for Clinical and Court subsets of the development set of third DIHARD challenge.}
  \vspace{-0.5cm}
  \begin{footnotesize}
  \begin{center}
  \label{Table:ResultsSubset}
  \centering
  \renewcommand{\arraystretch}{1.2}
  \begin{tabular}{|c|c|c|c|c|c|}
    \hline
    \multirow{2}{*}{Domain} & \multirow{2}{*}{Method} & \multicolumn{2}{|c|}{First Pass} & \multicolumn{2}{|c|}{Re-segmentation with VB-HMM}\\
    \cline{3-6}
       && Full & Core & Full & Core \\
    \hline
    \multirow{4}{*}{Clinical} & Baseline & 17.55 /	28.88&	16.08 /	26.38&	17.69 /	28.46&	16.72 /	27.21\\
     & Domain-dependent threshold and PLDA adaptation &20.06	/ 29.92	&18.88 /	28.11&	16.71	/	 25.50&	15.21 /	23.66\\
     & Domain-dependent threshold and PLDA adaptation with full-data & 15.81	/ 23.69	&14.61	 / 22.37&	14.69 /	 	22.07&	13.78 /	21.59\\
     & Same as above + domain-dependent parameter for PCA  & 14.67	/ 22.66	&12.91	/ 20.83	&13.79 /		20.65&	12.66 /	20.05\\
     \hline
    \multirow{4}{*}{Court}     & Baseline & 10.81 /	38.75&	10.81 /	38.75&	10.17 /		37.63&	10.17 /	37.63\\
     & Domain-dependent threshold and PLDA adaptation &12.19	/ 43.99 &	12.19 /	43.99 &	9.03 /	37.97 &		9.03 /	37.97\\
     & Domain-dependent threshold and PLDA adaptation with full-data & 5.82	/ 23.91 &	5.82 /	23.91 &	4.77 /	22.22 		&4.77 /	22.22\\
    & Same as above + domain-dependent parameter for PCA  & 5.03	/ 17.30 &	5.03 /	17.30 &	3.82 /	16.04 &		3.82 /	16.04\\ 
    \hline
  \end{tabular}
  \vspace{-0.8cm}
  \end{center}
  \end{footnotesize}
\end{table*}

\section{Acoustic domain identification system}
Our ADI system is based on the speaker embeddings as sentence-level feature and nearest neighbor classifier. Though the speaker embeddings are principally developed for speaker characterization, they also capture information related to acoustic scene~\cite{zeinali2018convolutional}, recording session~\cite{Probing2019}, and channel~\cite{Wang2017}. In this work, we study two frequently used speaker embeddings: discriminatively trained x-vectors and generative i-vectors. In an earlier study, these speaker embeddings were investigated for the second DIHARD dataset on related tasks~\cite{sahidullah2019speed,fennir2020acoustic}.  

The experiments were performed on the development set consisting of 254 speech utterances from 11 different domains. We randomly selected 200 utterances for training and used the remaining 54 for test. For similarity measurement between training and test data, \emph{cosine similarity} was employed. We repeated the experiments 1000 times and obtained average accuracy of 71.39\% and 90.81\% for x-vector and i-vector system, respectively. The comparison of ADI performance of the two systems, shown in Fig.~\ref{Fig:ADIaccuracy}, confirms that the i-vector system is substantially better than the x-vector system for ADI on DIHARD III dataset. In the subsequent experiments, we have used this system (but trained with all 254 sentences) for identifying the domains of evaluation data samples.

\section{Experimental setup}
\vspace{-0.1cm}
Our experimental setup is based on the baseline system created by the organizers~\cite{ryant2020thirdpaper}. We have used the toolkit\footnote{\url{https://github.com/dihardchallenge/dihard3_baseline}} with the same frame-level acoustic features, embedding extractor, scoring method, etc. In order to extract utterance-level embeddings for ADI task, we used pre-trained x-vector and i-vector model trained on VoxCeleb audio-data\footnote{\url{https://kaldi-asr.org/models/m7}}.

\section{Results on the challenge dataset}
\vspace{-0.1cm}

\subsection{Impact of domain-dependent processing}
In the first experiment with the baseline system, we analyzed the diarization performance for each domain and noticed that the performance varied with domains, as is expected. Next, the diarization experiment was separately performed with each domain. Interestingly, this degraded the performance compared to the baseline. This is more likely due to limited speaker and acoustic variability in domain-specific subset which is also used as in-domain target data for PLDA adaptation. Hence, we propose to apply domain-specific threshold for speaker clustering but PLDA adaptation is performed with audio-data from all the eleven subsets. The baseline system uses recording-dependent PCA during PLDA scoring where 30\% of the total energy is preserved during dimensionality reduction. We experimentally optimized this value for all the domains separately which further improved the diarization performance. In Table~\ref{Table:ResultsSubset}, we show the results for two subsets clinical and court. Table~\ref{Table:Dev} summarizes the results for the entire development set.

\begin{table}[h]
  \caption{Results showing the speaker diarization performance using baseline (B) and proposed methods (M1 and M2) on \textbf{development set} of third DIHARD challenge. M1: domain-dependent threshold, M2: domain-dependent threshold and domain-dependent parameter for PCA.}
  \vspace{-0.5cm}
  \begin{footnotesize}
  \begin{center}
  \label{Table:Dev}
  \centering
  \renewcommand{\arraystretch}{1.2}
  \begin{tabular}{|c|c|c|c|c|}
    \hline
    \multirow{2}{*}{Method} & \multicolumn{2}{|c|}{Full} & \multicolumn{2}{|c|}{Core}\\
     \cline{2-5}
      & DER & JER & DER & JER \\
    \hline
    B	& 19.59	&43.01	&20.17&	47.28\\
    M1	& 17.97	&40.33	&18.73	&44.77\\
    M2	& 17.40&	38.08&	17.95&	42.12\\
    \hline
  \end{tabular}
  \vspace{-0.5cm}
  \end{center}
  \end{footnotesize}
\vspace{-0.5cm}
\end{table}

\begin{table}[h]
  \caption{Same as Table ~\ref{Table:Dev} but for \textbf{evaluation set}.}
  \vspace{-0.5cm}
  \begin{footnotesize}
  \begin{center}
  \label{Table:Eval}
  \centering
  \renewcommand{\arraystretch}{1.2}
  \begin{tabular}{|c|c|c|c|c|}
    \hline
    \multirow{2}{*}{Method} & \multicolumn{2}{|c|}{Full} & \multicolumn{2}{|c|}{Core}\\
     \cline{2-5}
       & DER & JER & DER & JER \\
    \hline
    B (Submission ID: 1044)	& 19.19	&43.28&	20.39&	48.61\\
    M1 (Submission ID: 1218)	& 17.56	&38.60&	19.23&	43.74\\
    M2 (Submission ID: 1373)	& 17.20	&37.30&	18.66&	42.23\\
    \hline
  \end{tabular}
  \vspace{-0.5cm}
  \end{center}
  \end{footnotesize}
\vspace{-0.1cm}  
\end{table}

\subsection{Results on evaluation set}
We performed domain-dependent diarization on the evaluation set by first predicting the domain for every utterance, followed by grouping of the utterances according to the predicted domains. We used domain-specific threshold for speaker clustering and percentage of total energy for dimensionality reduction optimized on corresponding domain of the  development set. The results summarized in Table~\ref{Table:Eval} indicate substantial improvement over the baseline. 

\vspace{-0.3cm}
\section{Conclusion}
\vspace{-0.1cm}

In this work, we explored domain-dependent speaker diarization on the third DIHARD dataset by integrating an ADI system with the baseline system. Our ADI system is based on i-vector embeddings and nearest neighbour classifier. We applied domain-specific thresholds for speaker clustering and used domain-dependent PCA parameters for dimensionality reduction during PLDA scoring. The PLDA adaptation was performed with audio-data from all the domains. This way we achieved about ten percent relative improvement with respect to the baseline system for both the conditions in Track 1 of the challenge. The work can be extended with advanced embedding extractor based on ResNet, Extended-TDNN, etc.


\vspace{-0.3cm}

\section{Acknowledgements}
\vspace{-0.1cm}

\scriptsize Experiments presented in this paper were partially carried out using the Grid'5000 testbed, supported by a scientific interest group hosted by Inria and including CNRS, RENATER and several Universities as well as other organizations (see \url{https://www.grid5000.fr}).

\bibliographystyle{IEEEtran}

\bibliography{mybib}


\end{document}